\documentclass{llncs}
\usepackage{etex}
\usepackage[utf8]{inputenc}
\usepackage[T1]{fontenc}
\usepackage{ctable}

\newif\ifpublic
\publictrue

\usepackage{amsfonts,amsmath,amscd,amssymb,array}
\usepackage{mathtools}

\usepackage{type1cm}

\usepackage{fancyhdr}
\renewcommand{\headrulewidth}{0pt}
\usepackage[ruled,vlined,linesnumbered]{algorithm2e}
\SetEndCharOfAlgoLine{}

\usepackage{xcolor}
\definecolor{linkcolor}{rgb}{0.65,0,0}
\definecolor{citecolor}{rgb}{0,0.65,0}
\definecolor{urlcolor}{rgb}{0,0,0.65}
\usepackage[colorlinks=true, backref=page, linkcolor=linkcolor, urlcolor=urlcolor, citecolor=citecolor]{hyperref}
\usepackage{breakurl}

\usepackage{cite}
\usepackage{xspace}

\renewcommand{\tabcolsep}{4pt}

\def\F{\mathbb{F}}

\def\encode#1{\underline{\smash{#1}}}

\def\arraystretch{1.2}

\newcommand{\J}{\mathcal{J}}
\newcommand{\C}{\mathcal{C}}

\newcommand{\Mod}[1]{\text{ mod}\, #1}

\def\armcycles{$3\,589\,850$\xspace}
\def\armbytes{$7\,900$\xspace}
\def\armstack{$548$\xspace}

\def\armrom{$19\,606$\xspace}
\def\armscalrom{$4\,328$\xspace}
\def\armjacscalrom{$9\,874$\xspace}
\def\avrrom{$20\,242$\xspace}
\def\avrscalrom{$9\,490$\xspace}
\def\avrjacscalrom{$16\,516$\xspace}

\def\armbigintmul{$410$\xspace}
\def\armbigintsqr{$260$\xspace}
\def\armbigintred{$71$\xspace}
\def\armgfemul{$502$\xspace}
\def\armgfesqr{$353$\xspace}
\def\armgfemulconst{$83$\xspace}
\def\armgfeadd{$62$\xspace}
\def\armgfesub{$66$\xspace}

\def\armgfeinvert{$46\,091$\xspace}
\def\armgfesqrtinv{$48\,593$\xspace}
\def\armgfepowminhalf{$46\,294$\xspace}

\def\armadd{$62\,886$\xspace}
\def\armproject{$5\,667$\xspace}
\def\armxwrap{$49\,609$\xspace}
\def\armxunwrap{$2\,027$\xspace}
\def\armxaddnodiv{$9\,598$\xspace}
\def\armxdbladd{$9\,861$\xspace}
\def\armrecovergeneral{$88\,414$\xspace}
\def\armfastgenpartial{$6\,110$\xspace}
\def\armfastgenfull{$8\,333$\xspace}
\def\armrecoverfast{$124\,936$\xspace}
\def\armcompress{$2\,186$\xspace}
\def\armdecompress{$106\,013$\xspace}

\def\armscal{$2\,633\,662$\xspace}
\def\armjacscal{$2\,709\,401$\xspace}
\def\armjac2scal{$4\,270\,563$\xspace}

\def\armkeygen{$2\,774\,087$\xspace}
\def\armsign{$2\,865\,351$\xspace}
\def\armverify{$4\,453\,978$\xspace}

\def\armdhexchange{$2\,644\,604$\xspace}

\def\armkeygenram{$1\,056$\xspace}
\def\armsignram{$1\,360$\xspace}
\def\armverifyram{$1\,432$\xspace}

\def\armjacscalram{$968$\xspace}
\def\armdhexchangeram{$584$\xspace}
\def\armscalram{$248$\xspace}

\def\avrbigintmul{$1\,654$\xspace}
\def\avrbigintsqr{$1\,171$\xspace}
\def\avrbigintred{$438$\xspace}
\def\avrgfemul{$1\,952$\xspace}
\def\avrgfesqr{$\,1469$\xspace}
\def\avrgfemulconst{$569$\xspace}
\def\avrgfeadd{$400$\xspace}
\def\avrgfesub{$401$\xspace}

\def\avrgfeinvert{$169\,881$\xspace}
\def\avrgfesqrtinv{$178\,041$\xspace}
\def\avrgfepowminhalf{$169\,881$\xspace}

\def\avradd{$228\,552$\xspace}
\def\avrproject{$20\,205$\xspace}
\def\avrxwrap{$182\,251$\xspace}
\def\avrxunwrap{$7\,297$\xspace}
\def\avrxaddnodiv{$34\,774$\xspace}
\def\avrxdbladd{$36\,706$\xspace}
\def\avrrecovergeneral{$318\,910$\xspace}
\def\avrfastgenpartial{$21\,339$\xspace}
\def\avrfastgenfull{$29\,011$\xspace}
\def\avrrecoverfast{$447\,176$\xspace}
\def\avrcompress{$8\,016$\xspace}
\def\avrdecompress{$386\,524$\xspace}

\def\avrscal{$9\,513\,536$\xspace}
\def\avrjacscal{$9\,968\,127$\xspace}
\def\avrjac2scal{$15\,701\,276$\xspace}

\def\avrkeygen{$10\,206\,181$\xspace}
\def\avrsign{$10\,404\,033$\xspace}
\def\avrverify{$16\,240\,510$\xspace}

\def\avrdhexchange{$9\,739\,059$\xspace}

\def\avrkeygenram{$812$\xspace}
\def\avrsignram{$926$\xspace}
\def\avrverifyram{$992$\xspace}

\def\avrdhexchangeram{$429$\xspace}
\def\avrjacscalram{$735$\xspace}
\def\avrscalram{$99$\xspace}

\newcommand{\ie}{i.\,e.\ }

\newcommand{\Jac}[2][{}]{\ensuremath{\mathcal{J}_{#2}^{#1}}}
\newcommand{\Kum}[2][{}]{\ensuremath{\mathcal{K}_{#2}^{#1}}}
\newcommand{\Flynn}[2][{}]{\ensuremath{{\widetilde{\mathcal{K}}_{#2}}^{#1}}}
\newcommand{\subgrp}[1]{{\left\langle{#1}\right\rangle}}

\newcommand{\FF}{\mathbb{F}}
\newcommand{\PP}{\mathbb{P}}
\newcommand{\Mumford}[2]{{\left\langle{#1},{#2}\right\rangle}}
\newcommand{\PPsqr}{\mathcal{S}}
\newcommand{\PPmul}{\mathcal{M}}
\newcommand{\Hadamard}{\mathcal{H}}

\newcommand{\Kpt}[1]{\ensuremath{(x_{#1}:y_{#1}:z_{#1}:t_{#1})}}
\newcommand{\Fpt}[1]{\ensuremath{(\tilde x_{#1}:\tilde y_{#1}:\tilde z_{#1} : \tilde t_{#1})}}
\newcommand{\Fptshort}[1]{\ensuremath{(\tilde x_{#1}:\tilde y_{#1}:\tilde z_{#1})}}

\begin{document}

\titlerunning{\(\mu\)Kummer: hyperelliptic cryptography on microcontrollers}
\title{\(\mu\)Kummer: efficient hyperelliptic signatures\\
    and key exchange on microcontrollers}

\ifpublic
\author{
  Joost~Renes\inst{1}
\and
  Peter~Schwabe\inst{1}
\and
  Benjamin~Smith\inst{2} 
\and
  Lejla~Batina\inst{1}
\thanks{
This work has been supported 
by the Netherlands Organisation for Scientific Research (NWO) through Veni 2013 project 13114 and
by the Technology Foundation STW (project 13499 - TYPHOON \& ASPASIA), from the Dutch government.
Permanent ID of this document: {\tt b230ab9b9c664ec4aad0cea0bd6a6732}. 
Date: 2016-04-07} 
}
\institute{
  Digital Security Group, Radboud University, The Netherlands\\
  \email{lejla@cs.ru.nl,j.renes@cs.ru.nl,peter@cryptojedi.org}
  \and
  INRIA \emph{and} Laboratoire d'Informatique de l'École polytechnique (LIX), France\\
  \email{smith@lix.polytechnique.fr}
}
\fi 

\maketitle

\begin{abstract}
    We describe the design and implementation of
    efficient signature and key-exchange schemes
    for the AVR~ATmega and ARM Cortex~M0 microcontrollers,
    targeting the 128-bit security level.
    Our algorithms are based on an efficient Montgomery ladder scalar
    multiplication on the Kummer surface of Gaudry and Schost's 
    genus-2 hyperelliptic curve,
    combined with the Jacobian point recovery technique of Costello,
    Chung, and Smith.
    Our results are the first to show the feasibility of software-only
    hyperelliptic cryptography on constrained platforms, 
    and represent a significant improvement on the elliptic-curve 
    state-of-the-art for both key exchange and signatures on these architectures.
    Notably, our key-exchange scalar-multiplication software
    runs in under 9740k cycles on the ATmega, and under 2650k cycles on
    the Cortex M0.

\end{abstract}

\section{Introduction}
\label{sec:intro}

The current state of the art in asymmetric cryptography, not only on microcontrollers,
is elliptic-curve cryptography; the most widely accepted reasonable security is the
$128$-bit security level.
All current speed records for $128$-bit secure key exchange and
signatures on microcontrollers are held---until now---by elliptic-curve-based schemes.
Outside the world of microcontrollers,
it is well known that genus-2 hyperelliptic curves and their Kummer
surfaces
present an attractive alternative to elliptic curves.
For example, the current speed record for $128$-bit-secure scalar
multiplication on a range of architectures
is held by Kummer-based software presented at Asiacrypt 2014 by Bernstein, Chuengsatiansup, Lange, and Schwabe~\cite{BCLS14}.
These speed records were achieved by exploiting the computational power of 
vector units of recent ``large'' processors such as 
Intel Sandy Bridge, Ivy Bridge, and Haswell, or the ARM Cortex-A8.
Surprisingly, 
very little attention has been given to Kummer surfaces on embedded
processors.
Indeed, this is the first work showing the feasibility of software-only
implementations
of hyperelliptic-curve based crypto on constrained platforms.
There have been some investigations of binary hyperelliptic curves 
targeting the much lower 80-bit security level, 
but those are actually examples of software-hardware co-design 
showing that using hardware acceleration for field operations was
necessary to get reasonable performance figures
(see eg.~\cite{BatinaHHPV05} and~\cite{HodjatBHV07}).

In this paper we investigate the potential of genus-2 hyperelliptic curves
for both key exchange and signatures on the ``classical'' $8$-bit 
AVR ATmega architecture, and the more modern $32$-bit ARM Cortex-M0 processor.
We show that not only are hyperelliptic curves competitive, 
they clearly outperform
state-of-the art elliptic-curve schemes in terms of speed and size.
For example, our variable-basepoint scalar multiplication on a $127$-bit
Kummer surface is 31\% faster on AVR and 26\% faster on the M0 
than the recently presented speed records for Curve25519 software by
Düll, Haase, Hinterwälder, Hutter, Paar, Sánchez, and Schwabe~\cite{DHH+15};
our implementation is also smaller, and requires less RAM.

We use a recent result by Costello, Chung, and Smith~\cite{Recovery}
to also set new speed records for $128$-bit secure signatures.
Specifically, we present a new signature scheme based on fast
Kummer surface arithmetic. It is inspired by the EdDSA construction
by Bernstein, Duif, Lange, Schwabe, and Yang~\cite{EdDSA}. 
On the ATmega, it produces shorter signatures, achieves higher speeds 
and needs less RAM than the Ed25519 implementation presented in~\cite{NLD15}.

\begin{table}[h!]
\centering
\renewcommand{\tabcolsep}{0.1cm}
\renewcommand{\arraystretch}{1.1}
	\begin{tabular}{|c|c|c||c|c|}
\hline
& \multicolumn{2}{c||}{ATmega} & \multicolumn{2}{c|}{Cortex M0} \\
\hline
 & Cycles & Stack bytes & Cycles & Stack bytes \\
\hline
\hline
{\tt keygen} & \avrkeygen & \avrkeygenram & \armkeygen & \armkeygenram \\
\hline
{\tt sign} & \avrsign & \avrsignram & \armsign & \armsignram \\
\hline
{\tt verify} & \avrverify & \avrverifyram & \armverify & \armverifyram \\
\hline
{\tt dh\_exchange} & \avrdhexchange & \avrdhexchangeram & \armdhexchange & \armdhexchangeram \\
\hline
\end{tabular}
\vspace{0.2cm}
	\caption{Cycle counts and stack usage in bytes of all functions related to the signature and key exchange schemes, for the AVR ATmega and ARM Cortex M0 microcontrollers.}
	\label{tab:signaturesdh}
\end{table}

Our routines handling secret data are constant-time, 
and are thus naturally resistant to timing attacks.  
These algorithms are built around the Montgomery ladder, 
which improves resistance against simple-power-analysis
(SPA) attacks. 
Resistance to DPA attacks can be easily obtained
by randomizing the scalar and/or Jacobian points.
Re-randomizing the latter after each ladder step
also guarantees resistance against horizontal types of attacks.

\paragraph{Source code.}

We place all of the software described in this paper
into the public domain, to maximize reuseability of our results.
\ifpublic
The software is available at \url{http://www.cs.ru.nl/~jrenes/}.
\fi

\section{High-level overview}

We begin by describing 
the details of our signature and Diffie--Hellman schemes,
explaining the choices we made in their design.
Concrete details on their implementation
appear in~\S\ref{sec:building-blocks} and~\S\ref{sec:scalar-mult} below.
Experimental results and comparisons follow in~\S\ref{sec:results}.

\subsection{Signatures}\label{sec:highsig}

Our signature scheme adheres closely to the proposal of~\cite[\S8]{Recovery},
which in turn is a type of Schnorr signature~\cite{schnorr}. 
There are however some differences, and some possible trade-offs,
which we discuss below. The full scheme is presented at the end of this section. 

\paragraph{Group structure.}
We build the signature scheme on top of the group structure from the
Jacobian $\J_\C(\F_q)$ of a genus-2 hyperelliptic curve $\C$. 
More specifically, \(\C\) is the  
Gaudry--Schost curve over the prime field $\F_q$ with $q=2^{127}-1$
(cf.~\S\ref{sec:the-curve}). The Jacobian is a group of order
$\#\J_\C(\F_q)=2^4N$, where $$N=2^{250}-{\tt
0x334D69820C75294D2C27FC9F9A154FF47730B4B840C05BD}$$ is a 250-bit prime.
For more details on the Jacobian and its
elements, see~\S\ref{sec:jacobian-ops}.

\paragraph{Hash function.} 
The hash function $H$ can be any hash function 
with a 128-bit security level. 
For our purposes, $H(M)={\tt SHAKE128}(M, 512)$ suffices~\cite{FIPS202}.
While {\tt SHAKE128} has variable-length output, 
we shall only use the $512$-bit output implementation.

\paragraph{Encoding.} 
The objects on which we operate on the highest level are points $Q$ in
$\J_\C(\FF_q)$. To minimize communication costs, we compress the common
508-bit representation of $Q$ into 256 bits (see~\S\ref{sec:jacobian-ops}).
To avoid confusion between compressed and uncompressed points, 
we let $\encode{Q}$ denote the 256-bit encoding of~$Q$.
(This notation is the same as in~\cite{EdDSA}.)

\paragraph{Public generator.}
The public generator can be any element $P$ of $\J_\C(\FF_q)$ such that $[N]P=0$.
In our implementation we have made the arbitrary choice 
\( P = (X^2+u_1X+u_0, v_1X+v_0) \), where
\begin{align*}
    u_1 &= {\tt 0x7D5D9C3307E959BF27B8C76211D35E8A}, \\
    u_0 &= {\tt 0x2703150F9C594E0CA7E8302F93079CE8}, \\
    v_1 &= {\tt 0x444569AF177A9C1C721736D8F288C942}, \\
    v_0 &= {\tt 0x7F26CFB225F42417316836CFF8AEFB11}.
\end{align*}
This is the point which we use the most for scalar multiplication. 
Since it remains fixed, we assume we have its
decompressed representation precomputed, so as to avoid having to
perform the relatively expensive decompression operation whenever we need a 
scalar multiplication; this gives a low-cost speed gain. 
We further assume we have a ``wrapped'' representation of the
projection of \(P\) to the Kummer surface, which is used to speed
up the {\tt xDBLADD} function. 
See~\S\ref{sec:pseudo-mult} for more details on the {\tt xWRAP} function.

\paragraph{Public keys.}
In contrast to the public generator, we assume public keys are
compressed: they are communicated much more frequently,
and we therefore benefit much more from smaller keys. 
Moreover, we include the public key in one of the hashes during the {\tt sign} operation
\cite{KatzW03,MRaihiNPV98},
computing $h=H(\encode{R}||\encode{Q}||M)$ instead of 
the $h=H(\encode{R}||M)$ originally suggested by Schnorr~\cite{schnorr}.
This protects against adversaries
attacking multiple public keys simultaneously. 

\paragraph{Compressed signatures.}
Schnorr~\cite{schnorr}
mentions the option of compressing signatures
by hashing one of their two components:
the hash size only needs to be $b/2$ bits, where $b$ is the key length. 
Following this suggestion,
our signatures are 384-bit values of the form $(h_{128}||s)$,
where $h_{128}$ means the lowest 128 bits of $h=H(\encode{R}||\encode{Q}||M)$,
and $s$ is a 256-bit scalar. 
The most obvious upside is that signatures are smaller,
reducing communication overhead. 
Another big advantage is that we can exploit the half-size scalar to speed up signature verification.
On the other hand, we lose the possibility of efficient batch verification. 

\paragraph{Verification efficiency.} 
The most costly operation in signature verification is the
two-dimensional scalar multiplication $T=[s]P\oplus[h_{128}]Q$. 
In~\cite{Recovery}, the authors propose an algorithm relying on the
differential addition chains presented in~\cite{DJB-chain}.
However, since we are using compressed signatures, we have a small
scalar $h_{128}$. To abuse this, we simply compute $[s]P$ and
$[h_{128}]Q$ separately using the fast scalar multiplication on the
Kummer surface and finally add them together on the Jacobian. 
Not only do we need fewer cycles, but we can also reduce code size by
reusing the one-dimensional scalar multiplication routines. 

\paragraph{The scheme.}
We now define our signature scheme, taking the remarks above into account.

\begin{description}
    \item[Key generation ({\tt keygen}).]
        Let $d$ be a 256-bit secret key, 
        and $P$ the public generator. 
        Compute $(d'||d'') \gets H(d)$
        (with \(d'\) and \(d''\) both 256 bits),
        then $Q\gets [16d']P$. 
        The public key is~$\encode{Q}$.
    \item[Signing ({\tt sign}).]
        Let $M$ be a message, $d$ a 256-bit secret key, 
        $P$ the public generator,
        and $\encode{Q}$ a compressed public key. 
        Compute \((d'||d'') \gets H(d)\)
        (with \(d'\) and \(d''\) both 256 bits),
        then $r\gets H(d''||M)$, then $R\gets[r]P$,
        then $h\gets H(\encode{R}||\encode{Q}||M)$,
        and finally $s\gets\left(r-16h_{128}d'\right)\bmod{N}$.
        The signature is $(h_{128}||s)$.
    \item[Verification ({\tt verify}).]
        Let $M$ be a message with a signature $(h_{128}||s)$ 
        corresponding to a public key $\encode{Q}$,
        and let $P$ be the public generator.
        Compute $T\gets[s]P\oplus[h_{128}]Q$, 
        then $g\gets H(\encode{T}||\encode{Q}||M)$. 
        The signature is correct if $g_{128}=h_{128}$, 
        and incorrect otherwise.
\end{description}

\begin{remark}
    We note that there may be other, faster algorithms to compute this
    ``one-and-a-half-dimensional'' scalar multiplication. 
    Since for verification we do not have to worry about being constant-time, 
    one option might be to alter Montgomery's PRAC~\cite[\S3.3.1]{stamthesis} 
    to make use of the half-size scalar. 
    We have chosen not to pursue this line, 
    preferring the solid benefits of reduced code size instead.
\end{remark}

\subsection{Diffie-Hellman key exchange.}\label{sec:highdh}

For key exchange it is not necessary to have a group structure;
it is enough to have a pseudo-multiplication.
We can therefore carry out our the key exchange directly 
on the Kummer surface $\Kum{\C} = \Jac{\C}/\subgrp{\pm}$, 
gaining efficiency by not projecting from and recovering to
the Jacobian $\Jac{\C}$. 
If \(Q\) is a point on \(\Jac{\C}\),
then its image in \(\Kum{\C}\) is \(\pm Q\).
The common representation for points in \(\Kum{\C}(\FF_q)\)
is a 512-bit 4-tuple of field elements.
For input points (\ie the generator or public keys),
we prefer the 384-bit ``wrapped'' representation (see~\S\ref{sec:pseudo-add}).
This not only reduces key size, 
but it also allows a speed-up in the core {\tt xDBLADD} subroutine. 
The wrapped representation of a point \(\pm Q\) on \(\Kum{\C}\) 
is denoted by $\encode{\pm Q}$.

\begin{description}
    \item[Key exchange ({\tt dh\_exchange}).]
        Let $d$ be a 256-bit secret key, 
        and $\encode{\pm P}$ the public generator (respectively public key).
        Compute $\pm Q\gets\pm[d]P$. 
        The generated public key (respectively shared secret) is
        $\encode{\pm Q}$.
\end{description}

\begin{remark}
    While it might be possible to reduce the key size even further 
    to 256 bits, we would then have to pay the cost of compressing 
    and decompressing, and also wrapping for {\tt xDBLADD}
    (see the discussion in~\cite[App.~A]{Recovery}).
    We therefore choose to keep the 384-bit representation,
    which is consistent with~\cite{BCLS14}.
\end{remark} 

\section{
    Building blocks: algorithms and their implementation
}
\label{sec:building-blocks}

We begin by presenting the finite field \(\FF_{2^{127}-1}\)
in~\S\ref{sec:FF_q}.
We then define the curve \(\C\)
in~\S\ref{sec:the-curve},
before giving basic methods for the elements of \(\Jac{\C}\)
in~\S\ref{sec:jacobian-ops}.
We then present the fast Kummer \(\Kum{\C}\)
and its differential addition operations in~\S\ref{sec:fast-kummer}.

\subsection{The field \(\FF_q\)}
\label{sec:FF_q}

We work over the prime finite field $\F_q$, 
where \(q\) is the Mersenne prime
\[
    q := 2^{127} - 1 \ .
\]
We let \textbf{M}, \textbf{S}, \textbf{a}, \textbf{s}, and \textbf{neg}
denote the costs of multiplication, squaring, addition, subtraction, and
negation in \(\FF_q\).
Later, we will define a special operation for multiplying by small
constants: its cost is denoted by \(\mathbf{m_c}\).

We can represent elements of \(\FF_q\) as 127-bit values;
but since the ATmega and Cortex M0 work with 8- and 32-bit words,
respectively,
the obvious choice is to represent field elements with 128 bits.
That is, an element $g\in\FF_q$ is represented as $g=\sum_{i=0}^{15}g_i2^{8i}$ on the AVR ATmega platform and as $g=\sum_{i=0}^{3}g'_i2^{32i}$ on the Cortex M0, where $g_i\in\{0,\ldots,2^8-1\}$, $g'_i\in\{0,\ldots,2^{32}-1\}$. 

For complete field arithmetic we implement modular reduction, addition,
subtraction, multiplication, and inversion. 
We comment on some important aspects here, 
giving cycle counts in Table~\ref{tab:field}.

Working with the prime field $\F_q$, we need integer reduction modulo \(q\);
this is implemented as \texttt{bigint\_red}.
Reduction is very efficient because $2^{128}\equiv2\Mod{q}$, 
which enables us to reduce using only shifts and integer additions.  
Given this reduction,
we implement addition and subtraction operations for \(\FF_q\)
(as~\texttt{gfe\_add} and~\texttt{gfe\_sub}, respectively)
in the obvious way.

The most costly operations in \(\FF_q\)
are multiplication (\texttt{gfe\_mul})
and squaring (\texttt{gfe\_sqr}),
which are implemented as \(128\times128\)-bit bit integer operations
(\texttt{bigint\_mul} and \texttt{bigint\_sqrt})
followed by a call to \texttt{bigint\_red}.
Since we are working on the same platforms as in~\cite{DHH+15}
in which both of these operations are already highly optimized,
we chose to take the necessary code 
for \texttt{bigint\_mul} and \texttt{bigint\_sqr}
from those implementations:

\begin{itemize}
    \item
        On the AVR ATmega:
        The authors of~\cite{HS14} implement a 3-level
        Karatsuba multiplication of two 256-bit integers, representing
        elements $f$ of $\F_{2^{255}-19}$ as $f=\sum_{i=0}^{31}f_i
        2^{8i}$ with $f_i\in\{0,\ldots,2^8-1\}$. Since the first level
        of Karatsuba relies on a $128\times128$-bit integer
        multiplication routine named {\tt MUL128}, we simply lift this
        function out to form a 2-level $128\times128$-bit Karatsuba multiplication. Similarly, their $256\times256$-bit squaring relies on a $128\times128$-bit routine {\tt SQR128}, which we can (almost) directly use. Since the $256\times256$-bit squaring is 2-level Karatsuba, the $128\times128$-bit squaring is 1-level Karatsuba.
    \item
        On the ARM Cortex M0:
        The authors of~\cite{DHH+15} make use of optimized Karatsuba
        multiplication and squaring. In this case their assembly code
        does not rely on subroutines, but fully inlines 
        $128\times128$-bit multiplication and squaring. 
        The $256\times256$-bit multiplication and squaring are both 3-level
        Karatsuba implementations. Hence, using these, we end up with 
        2-level $128\times128$-bit Karatsuba multiplication and squaring.
\end{itemize}

The function \texttt{gfe\_invert}
computes inversions in \(\FF_q\) as exponentiations,
using the fact that \(g^{-1} = g^{q-2}\) for all \(g\) in \(\FF_q^\times\).
To do this efficiently we use an addition chain for $q-2$, doing the exponentiation in $10{\bf M}+126{\bf S}$.  

Finally, to speed up our Jacobian point decompression algorithms,
we define a function \texttt{gfe\_powminhalf}
which computes $g\mapsto g^{-1/2}$ for \(g\) in \(\FF_q\)
(up to a choice of sign). 
To do this, we note that 
\(
    g^{-1/2} = \pm g^{-(q+1)/4} =\pm g^{{(3q-5)}/{4}}
\) 
in~\(\FF_q\);
this exponentiation can be done with an addition chain of length 136, 
using $11{\bf M}+125{\bf S}$.
We can then define a function \texttt{gfe\_sqrtinv},
which given \((x,y)\) and a bit \(b\),
computes \((\sqrt{x},1/y)\) as \((\pm xyz,xyz^2)\)
where \(z = \mathtt{gfe\_powminhalf}(xy^2)\),
choosing the sign so that the square root has least significant bit \(b\).
Including the \texttt{gfe\_powminhalf} call,
this costs 15\textbf{M} + 126\textbf{S} + 1\textbf{neg}.

\begin{table}[h!]
\renewcommand{\tabcolsep}{0.1cm}
\renewcommand{\arraystretch}{1.1}
    \begin{center}
        \begin{tabular}{|c|c|c|c|}
            \hline
            & AVR ATmega & ARM Cortex M0 & Symbolic cost \\
            \hline
            \hline
            {\tt bigint\_mul} 
            & \avrbigintmul & \armbigintmul & \\
            \hline
            {\tt bigint\_sqr} 
            & \avrbigintsqr & \armbigintsqr & \\
            \hline
            {\tt bigint\_red} 
            & \avrbigintred & \armbigintred & \\
            \hline
            {\tt gfe\_mul} 
            & \avrgfemul & \armgfemul & \textbf{M} \\
            \hline
            {\tt gfe\_sqr} 
            & \avrgfesqr & \armgfesqr & \textbf{S} \\
            \hline
            {\tt gfe\_mulconst} 
            & \avrgfemulconst & \armgfemulconst & \(\mathbf{m_c}\) \\
            \hline
            {\tt gfe\_add} 
            & \avrgfeadd & \armgfeadd & \textbf{a} \\
            \hline
            {\tt gfe\_sub} 
            & \avrgfesub & \armgfesub & \textbf{s} \\
            \hline
            {\tt gfe\_invert} 
            & \avrgfeinvert & \armgfeinvert & \textbf{I} \\
            \hline
            {\tt gfe\_powminhalf} 
            & \avrgfepowminhalf & \armgfepowminhalf & 
            \(11\textbf{M} + 125\textbf{S}\)
            \\
            \hline
            {\tt gfe\_sqrtinv} 
            & \avrgfesqrtinv & \armgfesqrtinv & 
            \(15\textbf{M} + 126\textbf{S} + 1\textbf{neg}\)
            \\
            \hline
        \end{tabular}
    \end{center}
	\caption{
        Cycle counts for our field arithmetic implementation
        (including function-call overhead).
    }
	\label{tab:field}
\end{table}

\subsection{The curve \(\C\) and its theta constants} 
\label{sec:the-curve}

We define the curve \(\C\) ``backwards'',
starting from its (squared) theta constants
\[
    a := -11
    \ ,
    \quad
    b := 22
    \ ,
    \quad
    c := 19
    \ ,
    \quad
    \text{and}
    \quad
    d := 3
    \quad
    \text{in }
    \FF_q
    \ .
\]
From these, we define the dual theta constants
\begin{align*}
    A & {} := a + b + c + d = 33 \ ,
    &
    B & {} := a + b - c - d = -11 \ ,
    \\
    C & {} := a - b + c - d = -17 \ ,
    &
    D & {} := a - b - c + d = -49 \ .
\end{align*}
Observe that projectively,
\begin{align*}
    (\frac{1}{a}:\frac{1}{b}:\frac{1}{c}:\frac{1}{d}) 
    & = (114:-57:-66:-418)
    \shortintertext{and}
    (\frac{1}{A}:\frac{1}{B}:\frac{1}{C}:\frac{1}{D}) 
    & = (-833:2499:1617:561) \ .
\end{align*}
Crucially, all of these constants can be represented using just 16 bits.
Since Kummer arithmetic involves many multiplications by these constants,
we implement a separate $16\times128$-bit multiplication function {\tt
gfe\_mulconst}. For the AVR ATmega, we store the constants in
two 8-bit registers. For the Cortex M0, the values fit into a halfword;
this works well with the $16\!\times\!16$-bit multiplication. 
Multiplication by any of these 16-bit constants costs \(\mathbf{m_c}\).

Continuing, we define \(e/f := (1 + \alpha)/(1 - \alpha)\),
where \(\alpha^2 = CD/AB\) (we take the square root with least
significant bit 0), and thus
\begin{align*}
    \lambda := {ac}/{bd} 
    & = \mathtt{0x15555555555555555555555555555552}
    \ , 
    \quad
    \\
    \mu := {ce}/{df}
    & = \mathtt{0x73E334FBB315130E05A505C31919A746}
    \ , 
    \\
    \nu := {ae}/{bf}
    & = \mathtt{0x552AB1B63BF799716B5806482D2D21F3}
    \ .
\end{align*}
These are the \emph{Rosenhain invariants} of the curve \(\C\),
found by Gaudry and Schost~\cite{gaudry-schost},
which we are (finally!) ready to define as
\[
    \C : Y^2 = f_\C(X) := X(X-1)(X-\lambda)(X-\mu)(X-\nu)
    \ .
\]
The curve constants are the coefficients of \(f_\C(X) = \sum_{i=0}^5f_iX^i\):
so \(f_0 = 0\), \(f_5 = 1\), 
\begin{align*}
    f_1 & {} = \mathtt{0x1EDD6EE48E0C2F16F537CD791E4A8D6E} 
    \ ,
    \\
    f_2 & {} = \mathtt{0x73E799E36D9FCC210C9CD1B164C39A35} 
    \ ,
    \\
    f_3 & {} = \mathtt{0x4B9E333F48B6069CC47DC236188DF6E8}
    \ ,
    \\
    f_4 & {} = \mathtt{0x219CC3F8BB9DFE2B39AD9E9F6463E172}
    \ .
\end{align*}

We store the squared theta constants
\((a:b:c:d)\),
along with
\((1/a:1/b:1/c:1/d)\),
and \((1/A:1/B:1/C:1/D)\);
the Rosenhain invariants
\(\lambda\), \(\mu\), and \(\nu\),
together with \(\lambda\mu\) and \(\lambda\nu\);
and the curve constants \(f_1\), \(f_2\), \(f_3\), and \(f_4\),
for use in our Kummer and Jacobian arithmetic functions.
Obviously, none of the Rosenhain or curve constants are small;
multiplying by these costs a full \textbf{M}.

\subsection{Elements of \(\Jac{\C}\), compressed and decompressed.}
\label{sec:jacobian-ops}

%

Our algorithms use the usual Mumford representation
for elements of \(\Jac{\C}(\FF_q)\):
they correspond to pairs
\(
    \Mumford{u(X)}{v(X)}
\),
where \(u\) and \(v\) are polynomials over \(\FF_q\)
with \(u\) monic,
\(\deg v < \deg u \le 2\),
and
\(v(X)^2 \equiv f_\C(X) \pmod{u(X)}\).
We compute the group operation \(\oplus\) in \(\Jac{\C}(\FF_q)\)
using a function \texttt{ADD},
which implements the algorithm found in~\cite{jac-on-jac}
(after a change of coordinates to meet their Assumption~1)\footnote{We only call \texttt{ADD} once in our algorithms,
    so for lack of space we omit its description.}
at a cost of 
28\textbf{M} + 2\textbf{S} + 11\textbf{a} + 24\textbf{s} + 1\textbf{I}.

For transmission,
we compress the 508-bit Mumford representation 
to a 256-bit form.
Our functions
\texttt{compress} (Algorithm~\ref{alg:compress})
and \texttt{decompress} (Algorithm~\ref{alg:decompress})
implement Stahlke's compression technique (see~\cite{Stahlke04}
and~\cite[App.~A]{Recovery} for details).

\begin{algorithm}
    \caption{\texttt{compress}:
        compresses points on \(\Jac{\C}\) to 256-bit strings.
        Symbolic cost:
        3\textbf{M} + 1\textbf{S} + 2\textbf{a} + 2\textbf{s}.
        ATmega:
        \avrcompress{} cycles.
        Cortex M0:
        \armcompress{} cycles.
    }
    \label{alg:compress}
    \KwIn{\(\Mumford{X^2 + u_1X + u_0}{v_1X + v_0} = P \in \Jac{\C}\).}
    \KwOut{A string \(b_0\cdots b_{255}\) of 256 bits.}
    \SetKwData{b}{b}
    \SetKwData{w}{w}
    \SetKwFunction{LSB}{LeastSignificantBit}
    \(\w \gets 4((u_1\cdot v_0 - u_0\cdot v_1)\cdot v_1 - v_0^2)\)
    \tcp*{3\textbf{M} + 1\textbf{S} + 2\textbf{a} + 2\textbf{s}}
    \(\b_0 \gets \LSB(v_1)\)
    \;
    \(\b_{128} \gets \LSB(\w)\)
    \;
    \Return{\(\b_0||u_0||\b_{128}||u_1\)}
\end{algorithm}

\begin{algorithm}
    \caption{\texttt{decompress}:
        decompresses 256-bit string to a point on \(\Jac{\C}\).
        Symbolic cost:
        46\textbf{M}
        +
        255\textbf{S}
        + 
        17\textbf{a}
        +
        12\textbf{s}
        +
        6\textbf{neg}.
        ATmega:
        \avrdecompress{} cycles
        Cortex M0:
        \armdecompress{} cycles
    }
    \label{alg:decompress}
    \KwIn{A string \(b_0\cdots b_{255}\) of 256 bits.}
    \KwOut{\(\Mumford{X^2 + u_1X + u_0}{v_1X + v_0} = P \in \Jac{\C}\).}
    \SetKwData{T}{T}
    \SetKwData{U}{U}
    \SetKwData{V}{V}
    \SetKwFunction{LSB}{LeastSignificantBit}
    %
    \(\U_1 = b_{129}\cdots b_{256}\) as an element of \(\FF_q\)
    \;
    \(\U_0 = b_{1}\cdots b_{127}\) as an element of \(\FF_q\)
    \;
    %
    \(\T_1 \gets \U_1^2\)
    \tcp*{1S}
    \(\T_2 \gets \U_0 - \T_1\) 
    \tcp*{1s}
    \(\T_3 \gets \U_0 + \T_2\) 
    \tcp*{1a}
    \(\T_4 \gets \U_0\cdot(\T_3\cdot f_4 + (\U_1\cdot f_3 - 2f_2))\)
    \tcp*{3M + 1a + 2s}
    \(\T_3 \gets -\T_3\) 
    \tcp*{1neg}
    \(\T_1 \gets \T_3 - \U_0\) 
    \tcp*{1s}
    \(\T_4 \gets 2(\T_4 + (\T_1\cdot\U_0 + f_1) \cdot \U_1)\) 
    \tcp*{2M + 3a}
    \(\T_1 \gets 2(\T_1 - \U_0))\) 
    \tcp*{1a + 1s}
    \(\T_5 \gets ((\U_0 - (f_3 + \U_1\cdot(\U_1 - f_4)))\cdot\U_0 + f_1)^2\) 
    \tcp*{2M + 1S + 2a + 2s}
    \(\T_5 \gets \T_4^2 - 2\T_5\cdot\T_1\) 
    \tcp*{1M + 1S + 1a + 1s}
    \((\T_6,\T_5) \gets \mathtt{gfe\_sqrtinv}(\T_5,\T_1,b_1)\) 
    \tcp*{19M + 127S + 2neg}
    \(\T_4 \gets (\T_5 - \T_4)\cdot\T_6\) 
    \tcp*{1M + 1s}
    \(\T_5 \gets -f_4\cdot\T_2 - ((\T_3 - f_3)\cdot \U_1) + f_2 + \T_4\)
    \tcp*{2M + 2s + 2a + 1neg}
    \(\T_6 = \mathtt{gfe\_powminhalf}(4\T_6)\)
    \tcp*{\(=1/(2v_1)\). 11M + 125S + 2a}
    \(\V_1 \gets 2\T_5\cdot\T_6\)
    \tcp*{1M + 1a}
    \lIf(\tcp*[f]{2neg}){\(b_0 \not= \LSB(\V_1)\)}{
        \((\V_1,\T_6) \gets (-\V_1,-\T_6)\)
    }
    \(\T_5 \gets (\U_1\cdot f_4 + (\T_2 - f_3))\cdot\U_0\)
    \tcp*{2M + 1a + 1s}
    \(\V_0 \gets (\U_1\cdot\T_4 + \T_5 + f_1)\cdot\T_6\)
    \tcp*{2M + 2a}
    \Return{\(\Mumford{X^2 + \U_1 X + \U_0}{\V_1 X + \V_0}\)}
\end{algorithm}

\subsection{The Kummer surface \(\Kum{\C}\)}
\label{sec:fast-kummer}

The Kummer surface of \(\C\) 
is the quotient \(\Kum{\C}:= \Jac{\C}/\subgrp{\pm1}\);
points on \(\Kum{\C}\) correspond to points on \(\Jac{\C}\) taken up to
sign.
If \(P\) is a point in \(\Jac{\C}\),
then we write
\[
    (x_P:y_P:z_P:t_P) = \pm P
\]
for its image in \(\Kum{\C}\).
To avoid subscript explosion,
we make the following convention:
when points \(P\) and \(Q\) on \(\Jac{\C}\) are clear from the context,
we write
\[
    (x_\oplus:y_\oplus:z_\oplus:t_\oplus) = \pm(P\oplus Q)
    \quad
    \text{and}
    \quad
    (x_\ominus:y_\ominus:z_\ominus:t_\ominus) = \pm(P\ominus Q)
    \ .
\]

The Kummer surface of this \(\C\) has a ``fast'' model in~\(\PP^3\)
defined by
\[
    \Kum{\C}:
    E\cdot xyzt
    =
    \left(
        \begin{array}{c}
        (x^2 + y^2 + z^2 + t^2)
        \\
        - F\cdot (xt+yz) - G\cdot (xz+yt) - H\cdot (xy+zt)
        \end{array}
    \right)^2
\]
where
\[
    F = \frac{a^2-b^2-c^2+d^2}{ad-bc} \ ,
    \quad
    G = \frac{a^2-b^2+c^2-d^2}{ac-bd} \ ,
    \quad
    H = \frac{a^2+b^2-c^2-d^2}{ab-cd} \ ,
\]
and \(E = 4abcd\left(ABCD/((ad-bc)(ac-bd)(ab-cd))\right)^2 \)
(see eg.~\cite{Chudnovsky--Chudnovsky}, \cite{cosset}, and \cite{Gaudry}).
The identity point \(\Mumford{1}{0}\)
of \(\Jac{\C}\) maps to 
\[
    \pm 0_{\Jac{\C}} 
    = 
    ( a : b : c : d )
    \ .
\]
Algorithm~\ref{alg:projectFast}
(\texttt{Project})
projects general points from \(\Jac{\C}(\FF_q)\) into \(\Kum{\C}\).
The ``special'' case, where \(u\) is linear, is treated
in~\cite[\S7.2]{Recovery}.

\begin{algorithm}
    \caption{\texttt{Project}:
        \(\Jac{\C}\to\Kum{\C}\).
        Symbolic cost: 
        8\textbf{M} 
        + 1\textbf{S} 
        + 4\(\mathbf{m_c}\) 
        + 7\textbf{a} 
        + 4\textbf{s}. 
        ATmega:
        \avrproject{} cycles.
        Cortex M0:
        \armproject{} cycles.
    }
    \label{alg:projectFast}
    \KwIn{\(\Mumford{X^2 + u_1X + u_0}{v_1X + v_0} = P \in \Jac{\C}\).}
    \KwOut{\(\Kpt{P} = \pm P \in \Kum{\C}\).}
    \SetKwData{V}{V}
    \SetKwData{T}{T}
    \SetKwData{xP}{xP}
    \BlankLine
    %
    \(
        (\T_1,\T_2,\T_3,\T_4)
        \gets
        (\mu-u_0,\lambda\nu-u_0,\nu-u_0,\lambda\mu-u_0)
    \)
    \tcp*{4\textbf{s}}
    \(\T_5 \gets \lambda + u_1\)
    \tcp*{1a}
    \(\T_7 \gets u_0\cdot((\T_5 + \mu)\cdot\T_3)\)
    \tcp*{2M + 1a}
    \(\T_5 \gets u_0\cdot((\T_5 + \nu)\cdot\T_1)\)
    \tcp*{2M + 1a}
    \(
        (\T_6,\T_8) 
        \gets 
        (u_0\cdot((\mu + u_1) \cdot \T_2 + \T_2),
        u_0\cdot((\nu + u_1) \cdot \T_4 + \T_4))
    \)
    \tcp*{4M + 4a}
    %
    \(\T_1 \gets v_0^2\)
    \tcp*{1S}
    \(
        (\T_5,\T_6,\T_7,\T_8)
        \gets
        (\T_5-\T_1,\T_6-\T_1,\T_7-\T_1,\T_8-\T_1)
    \)
    \tcp*{4s}
    \Return{\((a\cdot\T_5:b\cdot\T_6:c\cdot\T_7:d\cdot\T_8)\)}
    \tcp*{4\(\mathbf{m_c}\)}
\end{algorithm}

\subsection{Pseudo-addition on \(\Kum{\C}\).}
\label{sec:pseudo-add}

While the points of \(\Kum{\C}\) do not form a group,
we have a pseudo-addition operation (differential addition),
which computes \(\pm(P\oplus Q)\)
from \(\pm P\), \(\pm Q\), and \(\pm (P\ominus Q)\).
The function \(\texttt{xADD}\) (Algorithm~\ref{alg:xADD})
implements the standard differential addition.
The special case where \(P = Q\)
yields a pseudo-doubling operation.

To simplify the presentation of our algorithms,
we define three operations on points in \(\PP^3\).
First, 
\(\PPmul: \PP^3\times\PP^3\to\PP^3\) multiplies the corresponding coordinates
of a pair of points:
\[
    \PPmul: 
    \left((x_1:y_1:z_1:t_1),(x_2:y_2:z_2:t_2)\right)
    \longmapsto
    (x_1x_2:y_1y_2:z_1z_2:t_1t_2)
    \ .
\]
The special case \((x_1:y_1:z_1:t_1) = (x_2:y_2:z_2:t_2)\) 
is denoted by
\[
    \PPsqr : (x:y:z:t) \longmapsto (x^2:y^2:z^2:t^2)
    \ .
\]
Finally,
the Hadamard transform\footnote{Observe that \((A:B:C:D) = \Hadamard((a:b:c:d))\)
    and, dually, \((a:b:c:d) = \Hadamard((A:B:C:D))\).}
is defined by
\[
    \Hadamard:
    (x:y:z:t) 
    \longmapsto 
    (x':y':z':t')
    \quad
    \text{where}
    \quad
    \left\lbrace
    \begin{array}{r@{\;=\;}l}
        x' & x + y + z + t \ ,
        \\
        y' & x + y - z - t \ ,
        \\
        z' & x - y + z - t \ ,
        \\
        t' & x - y - z + t \ .
    \end{array}
    \right.
\]
Clearly \(\PPmul\) and \(\PPsqr\), 
cost \(4\mathbf{M}\) and \(4\mathbf{S}\), 
respectively.
The Hadamard transform can easily be implemented with \(4\mathbf{a}+4\mathbf{s}\).
However, the additions and subtractions are relatively cheap, making function call overhead a large factor. 
To minimize this we inline the Hadamard transform, trading a bit of code size for efficiency.

\begin{algorithm}
    \caption{\texttt{xADD}:
        Differential addition
        on \(\Kum{\C}\).
        Symbolic cost: 
        \(
            14\mathbf{M} + 4\mathbf{S} + 4\mathbf{m_c} 
            + 12\mathbf{a} + 12\mathbf{s}
        \).
        ATmega:
        \avrxaddnodiv{} cycles.
        Cortex M0:
        \armxaddnodiv{} cycles.
    }
    \label{alg:xADD}
    \KwIn{\(
            (\pm P, \pm Q, \pm (P\ominus Q)) 
            \in \Kum[3]{\C}
        \)
        for some \(P\) and \(Q\) on \(\Jac{\C}\).
    }
    \KwOut{\( \pm (P\oplus Q) \in \Kum{\C}\).}
    \SetKwData{V}{V}
    \SetKwData{C}{C}
    \( (\V_1,\V_2) \gets (\Hadamard(\pm{P}),\Hadamard(\pm{Q})) \) 
    \tcp*{8\textbf{a} + 8\textbf{s}}
    \( \V_1 \gets \PPmul(\V_1,\V_2) \)
    \tcp*{4\textbf{M}}
    \( \V_1 \gets \PPmul(\V_1,(1/A:1/B:1/C:1/D)) \)
    \tcp*{4\(\mathbf{m_c}\)}
    \( \V_1 \gets \Hadamard(\V_1) \)
    \tcp*{4\textbf{a} + 4\textbf{s}}
    \( \V_1 \gets \PPsqr(\V_1) \)
    \tcp*{4\textbf{S}}
    \label{alg:xADD_nodiv:line6}
    \( 
        (\C_1,\C_2)
        \gets
        ( z_\ominus\cdot t_\ominus , x_\ominus\cdot y_\ominus )
    \)
    \tcp*{2\textbf{M}}
    \label{alg:xADD_nodiv:line7}
    \( 
        \V_2
        \gets
        \PPmul(
            (\C_1 : \C_1 : \C_2 : \C_2)
            ,
            (y_\ominus:x_\ominus:t_\ominus:z_\ominus)
        )
    \)
    \tcp*{4\textbf{M}}
    \Return{\( \PPmul(\V_1,\V_2) \)}
    \tcp*{4\textbf{M}}
\end{algorithm}

Lines~\ref{alg:xADD_nodiv:line6}
and~\ref{alg:xADD_nodiv:line7}
of Algorithm~\ref{alg:xADD}
only involve the third argument, \(\pm(P\ominus Q)\);
essentially, they compute the point
\((
    y_\ominus z_\ominus t_\ominus
    :
    x_\ominus z_\ominus t_\ominus
    :
    x_\ominus y_\ominus t_\ominus
    :
    x_\ominus y_\ominus z_\ominus 
)\)
(which is projectively equivalent to 
\((1/x_\ominus : 1/y_\ominus : 1/z_\ominus : 1/t_\ominus)\),
but requires no inversions; 
note that this is generally \emph{not} a point on \(\Kum{\C}\)).
In practice,
the pseudoadditions used in our scalar multiplication
all use a fixed third argument,
so it makes sense to precompute this ``inverted'' point
and to scale it by \(x_\ominus\) so that the first coordinate is \(1\),
thus saving \(7\textbf{M}\) in each subsequent differential addition
for a one-off cost of \(1\textbf{I}\).
The resulting data can be stored as the 3-tuple
\((x_\ominus/y_\ominus,x_\ominus/z_\ominus,x_\ominus/t_\ominus)\),
ignoring the trivial first coordinate:
this is the \emph{wrapped} form of \(\pm(P\ominus Q)\).
The function \texttt{xWRAP} (Algorithm~\ref{alg:xWRAP})
applies this transformation.

\begin{algorithm}
    \caption{\texttt{xWRAP}:
        \( (x:y:z:t) \mapsto (x/y,x/z,x/t) \).
        Symbolic cost: 
        7\textbf{M} + 1\textbf{I}
        ATmega:
        \avrxwrap{} cycles.
        Cortex M0:
        \armxwrap{} cycles.
    }
    \label{alg:xWRAP}
    \KwIn{\((x:y:z:t) \in \PP^3\)}
    \KwOut{\((x/y,x/z,x/t) \in \FF_q^3\).}
    \SetKwData{V}{V}
    \(\V_1 \gets y\cdot z\)
    \tcp*{1M}
    \(\V_2 \gets x/(\V_1\cdot t) \)
    \tcp*{2M + 1I}
    \(\V_3 \gets \V_2\cdot t\)
    \tcp*{1M}
    \Return{\((\V_3\cdot z, \V_3\cdot y, \V_1\cdot\V_2)\)}
    \tcp*{3M}
\end{algorithm}

Algorithm~\ref{alg:xDBLADD}
combines the pseudo-doubling with the differential addition,
sharing intermediate operands,
to define a differential double-and-add \(\texttt{xDBLADD}\).
This is the fundamental building block of the Montgomery ladder.

\begin{algorithm}
    \caption{\texttt{xDBLADD}: 
        Combined differential double-and-add.
        The difference point is wrapped.
        Symbolic cost:
        \(
            7\mathbf{M} + 12\mathbf{S} + 12\mathbf{m_c}
            + 16\mathbf{a} + 16\mathbf{s}
        \).
        ATmega:
        \avrxdbladd{} cycles.
        Cortex M0:
        \armxdbladd{} cycles.
    }
    \label{alg:xDBLADD}
    \KwIn{\( (\pm P, \pm Q, (x_\ominus/y_\ominus,x_\ominus/z_\ominus,x_\ominus/t_\ominus)) \in \Kum[2]{\C}\times\FF_q\).}
    \KwOut{\( (\pm[2]P, \pm (P\oplus Q)) \in \Kum[2]{\C} \).}
    \SetKwData{V}{V}
    \( (\V_1,\V_2) \gets (\PPsqr(\pm{P}),\PPsqr(\pm{Q})) \) 
    \tcp*{8\textbf{S}}
    \( (\V_1,\V_2) \gets (\Hadamard(\V_1),\Hadamard(\V_2)) \) 
    \tcp*{8\textbf{a} + 8\textbf{s}}
    \( (\V_1,\V_2) \gets (\PPsqr(\V_1),\PPmul(\V_1,\V_2)) \)
    \tcp*{4\textbf{M}+4\textbf{S}}
    \( 
        (\V_1,\V_2) 
        \gets 
        \left(\PPmul(\V_1,(\frac{1}{A}:\frac{1}{B}:\frac{1}{C}:\frac{1}{D})),
        \PPmul(\V_2,(\frac{1}{A}:\frac{1}{B}:\frac{1}{C}:\frac{1}{D}))\right)
    \)
    \tcp*{8\(\mathbf{m_c}\)}
    \( (\V_1,\V_2) \gets (\Hadamard(\V_1),\Hadamard(\V_2)) \)
    \tcp*{8\textbf{a} + 8\textbf{s}}
    \Return{\( 
        (
        \PPmul(\V_1, (\frac{1}{a}:\frac{1}{b}:\frac{1}{c}:\frac{1}{d})),
            \PPmul(\V_2,
                (1:\frac{x_\ominus}{y_\ominus}:\frac{x_\ominus}{y_\ominus}:
                \frac{x_\ominus}{t_\ominus})
            )
        )
    \)}
    \tcp*{3\textbf{M} + 4\(\mathbf{m}_c\)}
\end{algorithm}

\begin{table}[h!]
\centering
\renewcommand{\tabcolsep}{0.1cm}
\renewcommand{\arraystretch}{1.1}
	\begin{tabular}{|c|c|c|c|c|c|c|c|r|r|}
\hline
 & ${\bf M}$ & ${\bf S}$ & ${\bf m_c}$ & ${\bf a}$ & ${\bf s}$ & ${\bf neg}$ & ${\bf I}$ & ATmega & Cortex M0\\
\hline
\hline
{\tt ADD} & 28 & 2 & 0 & 11 & 24 & 0 & 1 & \avradd & \armadd \\
\hline
{\tt Project} & 8 & 1 & 4 & 7 & 8 & 0 & 0 & \avrproject & \armproject \\
\hline
{\tt xWRAP} & 7 & 0 & 0 & 0 & 0 & 0 & 1 & \avrxwrap & \armxwrap \\
\hline
{\tt xUNWRAP} & 4 & 0 & 0 & 0 & 0 & 0 & 0 & \avrxunwrap & \armxunwrap \\
\hline
{\tt xADD} & 14 & 4 & 4 & 12 & 12 & 0 & 0 & \avrxaddnodiv & \armxaddnodiv \\
\hline
{\tt xDBLADD} & 7 & 12 & 12 & 16 & 16 & 0 & 0 & \avrxdbladd & \armxdbladd \\
\hline
{\tt recoverGeneral} & 77 & 8 & 0 & 19 & 10 & 3 & 1 & \avrrecovergeneral & \armrecovergeneral \\
\hline
{\tt fast2genPartial} & 11 & 0 & 0 & 9 & 0 & 0 & 0 & \avrfastgenpartial & \armfastgenpartial \\
\hline
{\tt fast2genFull} & 15 & 0 & 0 & 12 & 0 & 0 & 0 & \avrfastgenfull & \armfastgenfull \\
\hline
{\tt recoverFast} & 139 & 12 & 4 & 70 & 22 & 5 & 1 & \avrrecoverfast & \armrecoverfast \\
\hline
{\tt compress} & 3 & 1 & 0 & 2 & 2 & 0 & 0 & \avrcompress & \armcompress \\
\hline
{\tt decompress} & 46 & 255 & 0 & 17 & 12 & 6 & 0 & \avrdecompress & \armdecompress\\
\hline
\end{tabular}
\vspace{0.2cm}
	\caption{Operation and cycle counts of basic functions on the Kummer and Jacobian.}
	\label{tab:functions}
\end{table}

\section{
    Scalar multiplication
}
\label{sec:scalar-mult}

All of our cryptographic routines are built around
scalar multiplication in \(\Jac{\C}\) 
and pseudo-scalar multiplication in \(\Kum{\C}\).
We implement pseudo-scalar multiplication
using the classic Montgomery ladder in~\S\ref{sec:pseudo-mult}.
In~\S\ref{sec:recovery},
we extend this to full scalar multiplication on \(\Jac{\C}\)
using the point recovery technique proposed in~\cite{Recovery}.

\subsection{Pseudomultiplication on \(\Kum{\C}\)}
\label{sec:pseudo-mult}

Since \([m](\ominus P) = \ominus[m]P\) for all \(m\) and \(P\),
we have a pseudo-scalar multiplication operation
\((m,\pm P)\longmapsto \pm[m]P\) on \(\Kum{\C}\),
which we compute using Algorithm~\ref{alg:Ladder} 
(the Montgomery ladder), implemented as \texttt{crypto\_scalarmult}.
The loop of Algorithm~\ref{alg:Ladder}
maintains the following invariant:
at the end of iteration \(i\)
we have 
\[
    (V_1,V_2) = (\pm[k]P,\pm[k+1]P)  
    \quad
    \text{where}
    \quad
    \textstyle
    k = \sum_{j=i}^{\beta-1}m_j2^{\beta-1-i}
    \ .
\]
Hence, at the end we return \(\pm[m]P\), 
and also \(\pm[m+1]P\) as a (free) byproduct.
We assume that we have a constant-time conditional swap routine
\(\texttt{CSWAP}(b,(V_1,V_2))\),
which returns \((V_1,V_2)\) if \(b = 0\)
and \((V_2,V_1)\) if \(b = 1\).
This makes the execution of Algorithm~\ref{alg:Ladder}
uniform and constant-time,
which means it is suitable for use with secret values of \(m\).

\begin{algorithm}
    \caption{\texttt{crypto\_scalarmult}: 
        Montgomery ladder on \(\Kum{\C}\).
        Uniform and constant-time:
        may be used for secret scalars.
        The point is wrapped.
        Symbolic cost:
        \(
            (4 + 7\beta)\mathbf{M}      
            +
            12\beta\mathbf{S}      
            +
            12\beta\mathbf{m_c}    
            +
            16\beta\mathbf{a} 
            +
            16\beta\mathbf{s} 
        \),
        where \(\beta =\) scalar bitlength.
        ATmega:
        {\avrscal} cycles.
        Cortex:
        {\armscal} cycles.
    }
    \label{alg:Ladder}
    \KwIn{\( (m = \sum_{i=0}^{\beta-1}m_i2^i,(x_P/y_P,x_P/z_P,x_P/t_P))
        \in [0,2^\beta)\times\FF_q^3 
        \)
        for \(\pm P\) in \(\Kum{\C}\).
    }
    \KwOut{\( (\pm [m]P,\pm[m+1]P) \in \Kum[2]{\C} \).}
    \SetKwData{V}{V}
    \( \V_1 \gets (a:b:c:d) \)
    \;
    \( \V_2 \gets \texttt{xUNWRAP}(x_P/y_P,x_P/z_P,x_P/t_P) \)
    \tcp*{\(=\pm P\). 4M}
    \For(\tcp*[f]{7\(\beta\)M + 12\(\beta\)S + 12\(\beta m_c\) +
    16\(\beta\)a + 16\(\beta\)s}
){\(i = 250\) down to \(0\)}{
        \((\V_1,\V_2)\gets\texttt{CSWAP}(m_i,(\V_1,\V_2))\)
        \;
        \(
            (\V_1,\V_2)
            \gets 
            \texttt{xDBLADD}(\V_1,\V_2,(x_P/y_P,x_P/z_P,x_P/t_P))
        \)
        \;
        \((\V_1,\V_2)\gets\texttt{CSWAP}(m_i,(\V_1,\V_2))\)
    }
    \Return{ \( (\V_1,\V_2) \) }
\end{algorithm}

Our implementation of \texttt{crypto\_scalarmult} 
assumes that its input Kummer point \(\pm P\) is wrapped.
This follows the approach of~\cite{BCLS14}.
Indeed, many calls to \texttt{crypto\_scalarmult}
involve Kummer points that are stored or transmitted in wrapped form.
However, \texttt{crypto\_scalarmult} does require
the unwrapped point internally---if only to initialize one variable.
We therefore define a function \texttt{xUNWRAP} (Algorithm~\ref{alg:xUNWRAP})
to invert the \texttt{xWRAP} transformation
at a cost of only 4\textbf{M}.

\begin{algorithm}
    \caption{\texttt{xUNWRAP}:
        \( (x/y,x/z,x/t) \mapsto (x:y:z:t) \).
        Symbolic cost: 4\textbf{M}.
        ATmega:
        \avrxunwrap{} cycles.
        Cortex:
        \armxunwrap{} cycles.
    }
    \label{alg:xUNWRAP}
    \KwIn{\((u,v,w)\in \FF_q^3\) 
        s.t. \(u = x_P/y_P, v = x_P/z_P, w = x_P/t_P\) 
        for \(\pm P \in \Kum{\C}\)}
    \KwOut{\((x_P:y_P:z_P:t_P) \in \PP^3\)}
    \SetKwData{V}{T}
    \(
        (\V_1,\V_2,\V_3) 
        \gets 
        (v\cdot w, u\cdot w, u\cdot v)
    \)
    \tcp*{3\textbf{M}}
    \Return{\( (\V_3\cdot w:\V_1:\V_2:\V_3) \)}
    \tcp*{1\textbf{M}}
\end{algorithm}

\subsection{Point recovery from \(\Kum{\C}\) to \(\Jac{\C}\)}
\label{sec:recovery}

Point recovery means efficiently computing \([m]P\) on \(\Jac{\C}\)
given \(\pm[m]P\) on \(\Kum{\C}\)
and some additional information.
In our case, the additional information
is the base point \(P\)
and the second output of the Montgomery ladder,
\(\pm[m+1]P\).

Algorithm~\ref{alg:recoverFast} (\texttt{Recover}) implements
the point recovery algorithm described in~\cite{Recovery}.
This is the genus-2 analogue of the point recovery methods
defined for elliptic curves in~\cite{Lopez--Dahab}, \cite{Okeya--Sakurai},
and~\cite{Brier--Joye}.

\begin{algorithm}
    \caption{\texttt{Recover}: 
        From \(\Kum{\C}\) to \(\Jac{\C}\).
        Symbolic cost:
        139\textbf{M}
        +
        12\textbf{S}
        +
        4\(\mathbf{m_c}\)
        +
        70\textbf{a}
        +
        22\textbf{s}
        +
        3\textbf{neg}
        +
        1\textbf{I}.
        ATmega:
        \avrrecoverfast{} cycles.
        Cortex:
        \armrecoverfast{} cycles.
    }
    \label{alg:recoverFast}
    \KwIn{\( (P,\pm P,\pm Q,\pm(P\oplus Q))\in\Jac{\C}\times\Kum[3]{\C}\)
        for some \(P, Q\) in~\(\Jac{\C}\).
    }
    \KwOut{\(Q\in\Jac{\C}\). }
    \SetKwData{xD}{xD}
    \SetKwData{gP}{gP}
    \SetKwData{gS}{gS}
    \SetKwData{gD}{gD}
    \SetKwData{gQ}{gQ}
    \SetKwData{Q}{Q}
    \SetKwFunction{recoverGeneral}{recoverGeneral}
    \( \gP \gets \texttt{fast2genPartial}(\pm{P}) \)
    \tcp*{11M + 9a}
    \( \gQ \gets \texttt{fast2genFull}(\pm{Q}) \)
    \tcp*{15M + 12a}
    \( \gS \gets\texttt{fast2genPartial}(\pm(P\oplus Q))\)
    \tcp*{11M + 9a}
    \( 
        \xD
        \gets
        \texttt{xADD}(\pm P, \pm Q,\pm (P\oplus Q))
    \)
    \tcp*{14M + 4S + 4m\_c + 12a + 12s}
    \( \gD \gets\texttt{fast2genPartial}(\xD)\)
    \tcp*{11{M} + 9{a}}
    \Return{\(\recoverGeneral( P,\gP,\gQ,\gS,\gD)\)}
    \tcp*{77{M}+8{S}+19{a}+10{s}+3{neg}+1{I}}
\end{algorithm}

While we refer the reader to~\cite{Recovery}
for technical details on this method, and proof of its correctness,
there is one important mathematical detail that we should mention,
since it is reflected in the structure of our code.
Namely, point recovery is more naturally computed
starting from the general Flynn model \(\Flynn{\C}\) of the Kummer,
because it is more closely related to the Mumford model for \(\Jac{\C}\).
Algorithm~\ref{alg:recoverFast} therefore
proceeds in two steps:
first we map the problem onto \(\Flynn{\C}\)
using Algorithms~\ref{alg:fast2genFull}
and~\ref{alg:fast2genPartial}
(\texttt{fast2genFull} and~\texttt{fast2genPartial}),
and then we recover from \(\Flynn{\C}\) to \(\Jac{\C}\)
using Algorithm~\ref{alg:recoverGeneral} (\texttt{recoverGeneral}).

Since the general Kummer \(\Flynn{\C}\) 
only appears briefly in our recovery procedure 
(we never use its relatively slow arithmetic operations),
we will not investigate it any further here---but the curious reader may refer to~\cite{casselsflynn}
for the general theory.
For our purposes, it suffices to recall that 
\(\Flynn{\C}\) is,
like \(\Kum{\C}\), 
embedded in~\(\PP^3\);
and the isomorphism
\(\Kum{\C}\to\Flynn{\C}\)
is defined (in eg.~\cite[\S7.4]{Recovery}) by the linear transformation
\[
    \Kpt{P} \longmapsto \Fpt{P} := \Kpt{P}L 
    \ ,
\]
where \(L\) is (any scalar multiple of) the matrix
\[
    \left(
        \begin{array}{c@{\ \ }c@{\ \ }c@{\ \ }c}
            a^{-1}(\nu - \lambda)
            &
            a^{-1}(\mu\nu - \lambda)
            &
            a^{-1}\lambda\nu(\mu - 1)
            &
            a^{-1}\lambda\nu(\mu\nu - \lambda)
            \\
            b^{-1}(\mu - 1)
            &
            b^{-1}(\mu\nu - \lambda)
            &
            b^{-1}\mu(\nu - \lambda)
            &
            b^{-1}\mu(\mu\nu - \lambda)
            \\
            c^{-1}(\lambda - \mu)
            &
            c^{-1}(\lambda-\mu\nu)
            &
            c^{-1}\lambda\mu(1 - \nu)
            &
            c^{-1}\lambda\mu(\lambda-\mu\nu)
            \\
            d^{-1}(1-\nu)
            &
            d^{-1}(\lambda-\mu\nu)
            &
            d^{-1}\nu(\lambda-\mu)
            &
            d^{-1}\nu(\lambda-\mu\nu)
        \end{array}
    \right)
    \ ,
\]
which we precompute and store.
If \(\pm P\) is a point on \(\Kum{\C}\),
then \(\widetilde{\pm P}\) 
denotes its image on~\(\Flynn{\C}\);
we compute \(\widetilde{\pm P}\)
using Algorithm~\ref{alg:fast2genFull} (\texttt{fast2genFull}).

\begin{algorithm}
    \caption{\texttt{fast2genFull}: 
        The map \(\Kum{\C}\to\Flynn{\C}\).
        Symbolic cost: 
        \(15\mathbf{M}+12\mathbf{a}\).
        ATmega:
        \avrfastgenfull{} cycles.
        Cortex:
        \armfastgenfull{} cycles.
    }
    \label{alg:fast2genFull}
    \KwIn{\(\pm P\in\Kum{\C}\)}
    \KwOut{\(\widetilde{\pm P}\in\Flynn{\C}\).}
    \(
        \tilde x_P 
        \gets 
        x_P 
        + (L_{12}/L_{11})y_P 
        + (L_{13}/L_{11})z_P
        + (L_{14}/L_{11})t_P
    \)
    \tcp*{\(3\mathbf{M} + 3\mathbf{a}\)}
    \(
        \tilde y_P 
        \gets 
        (L_{21}/L_{11})x_P 
        + (L_{22}/L_{11})y_P
        + (L_{23}/L_{11})z_P
        + (L_{24}/L_{11})t_P
    \)
    \tcp*{\(4\mathbf{M} + 3\mathbf{a}\)}
    \(
        \tilde z_P 
        \gets
        (L_{31}/L_{11})x_P
        + (L_{32}/L_{11})y_P
        + (L_{33}/L_{11})z_P
        + (L_{34}/L_{11})t_P
    \)
    \tcp*{\(4\mathbf{M} + 3\mathbf{a}\)}
    \(
        \tilde t_P
        \gets
        (L_{41}/L_{11})x_P
        + (L_{42}/L_{11})y_P
        + (L_{43}/L_{11})z_P
        + (L_{44}/L_{11})t_P
    \)
    \tcp*{\(4\mathbf{M} + 3\mathbf{a}\)}
    \Return{ \(\Fpt{P}\) }
\end{algorithm}

Sometimes we only require the first three coordinates
of \(\widetilde{\pm P}\).
Algorithm~\ref{alg:fast2genPartial} (\texttt{fast2genPartial}) 
saves \(4\textbf{M}+3\textbf{a}\) per point by not computing~\(\tilde{t}_P\).

\begin{algorithm}
    \caption{\texttt{fast2genPartial}: 
        The map \(\Kum{\C}\to\PP^2\).
        Symbolic cost: \(11\mathbf{M}+9\mathbf{a}\).
        ATmega:
        \avrfastgenpartial{} cycles.
        Cortex:
        \armfastgenfull{} cycles.
    }
    \label{alg:fast2genPartial}
    \KwIn{\(\pm P \in \Kum{\C}\).}
    \KwOut{\(\Fptshort{P}\in\PP^2\)}
    \(
        \tilde x_P 
        \gets 
        x_P 
        + (L_{12}/L_{11})y_P 
        + (L_{13}/L_{11})z_P
        + (L_{14}/L_{11})t_P
    \)
    \tcp*{\(3\mathbf{M} + 3\mathbf{a}\)}
    \(
        \tilde y_P 
        \gets 
        (L_{21}/L_{11})x_P 
        + (L_{22}/L_{11})y_P
        + (L_{23}/L_{11})z_P
        + (L_{24}/L_{11})t_P
    \)
    \tcp*{\(4\mathbf{M} + 3\mathbf{a}\)}
    \(
        \tilde z_P 
        \gets
        (L_{31}/L_{11})x_P
        + (L_{32}/L_{11})y_P
        + (L_{33}/L_{11})z_P
        + (L_{34}/L_{11})t_P
    \)
    \tcp*{\(4\mathbf{M} + 3\mathbf{a}\)}
    \Return{ \(\Fptshort{P}\) }
\end{algorithm}

\begin{algorithm}
    \caption{\texttt{recoverGeneral}: 
        From \(\Flynn{C}\) to \(\Jac{\C}\).
        Symbolic cost:
        \(
            77\mathbf{M} 
            + 8\mathbf{S} 
            + 19\mathbf{a} 
            + 10\mathbf{s} 
            + 3\mathbf{neg}
            + 1\mathbf{I} 
        \).
        ATmega:
        \avrrecovergeneral{} cycles.
        Cortex:
        \armrecovergeneral{} cycles.
    }
    \label{alg:recoverGeneral}
    \KwIn{\(
            (P,
            \widetilde{\pm P},
            \widetilde{\pm Q},
            \widetilde{\pm(P\!\oplus\!Q)},
            \widetilde{\pm(P\!\ominus\!Q)}
            )
            \in
            \Jac{\C}\times\Flynn[4]{\C}
            \)
            for some \(P\) and \(Q\) in \(\Jac{\C}\).
        \\
        \qquad
        The values of \(\tilde t_P\), \(\tilde t_\oplus\),
        and \(\tilde t_\ominus\)
        are not required.
    }
    \KwOut{
        \(Q\in\Jac{\C}.\) 
    }
    \SetKwData{Zone}{Z1}
    \SetKwData{Ztwo}{Z2}
    \SetKwData{mZthree}{mZ3}
    \SetKwData{Zfour}{Z4}
    \SetKwData{D}{D}
    \SetKwData{E}{E}
    \SetKwData{F}{F}
    \SetKwData{Fi}{Fi}
    \SetKwData{Xone}{X1}
    \SetKwData{Xtwo}{X2}
    \SetKwData{Xthree}{X3}
    \SetKwData{Xfour}{X4}
    \SetKwData{Xfive}{X5}
    \SetKwData{Xsix}{X6}
    \SetKwData{Xseven}{X7}
    \SetKwData{Xeight}{X8}
    \SetKwData{Xnine}{X9}
    \SetKwData{Xten}{X10}
    \SetKwData{Cfive}{C5}
    \SetKwData{Csix}{C6}
    \SetKwData{Tone}{T1}
    \SetKwData{Ttwo}{T2}
    \SetKwData{Tthree}{T3}
    \SetKwData{Tseven}{T4} 
    \SetKwData{Tthirteen}{T5} 
    \SetKwData{Tsix}{T6}
    \SetKwData{Uzero}{U0}
    \SetKwData{Uone}{U1}
    \( 
        (\Zone,\Ztwo)
        \gets 
        (
            \tilde y_P\cdot \tilde x_Q - \tilde x_Q\cdot \tilde y_P 
            ,
            \tilde x_P\cdot \tilde z_Q - \tilde z_P\cdot \tilde x_Q 
        )
    \)
    \tcp*{4\textbf{M}+2\textbf{s}}
    \( \Tone \gets \Zone\cdot \tilde z_P \)
    \tcp*{1\textbf{M}}
    % ...
    \( \mZthree \gets \Ztwo\cdot\tilde y_P + \Tone \)
    \tcp*{1\textbf{M} + 1\textbf{a}}
    % ...
    \( \D \gets \Ztwo^2\cdot \tilde x_P + \mZthree\cdot\Zone \)
    \tcp*{2\textbf{M} + 1\textbf{S} + 1\textbf{a}}
    \( \Ttwo \gets \Zone\cdot\Ztwo \)
    \tcp*{1\textbf{M}}
    % ...
    \( \Tthree \gets \tilde x_P\cdot \tilde x_Q \)
    \tcp*{1\textbf{M}}
    % ...
    \( 
        \E 
        \gets 
        \Tthree
        \cdot(
            \Tthree\cdot (f_2\cdot \Ztwo^2 - f_1\cdot\Ttwo)
            +
            \tilde t_Q\cdot \D
        )
    \)
    \tcp*{5\textbf{M} + 1\textbf{S} + 1\textbf{a} + 1\textbf{s}}
    % ...
    \( 
        \E 
        \gets 
        \E
        +
        \mZthree\cdot\tilde{x}_Q^2 
        \cdot 
        (f_3\cdot \Ztwo\cdot \tilde x_P + f_4\cdot\mZthree)
    \)
    \tcp*{5\textbf{M} + 1\textbf{S} + 2\textbf{a}}
    % ...
    \(
        \E 
        \gets
        \E
        +
        \mZthree
        \cdot
        \tilde x_Q
        \cdot (
            \mZthree\cdot\tilde y_Q - \Ztwo\cdot \tilde x_P\cdot\tilde z_Q
        )
    \)
    \tcp*{5\textbf{M} + 1\textbf{a} + 1\textbf{s}}
    \( \Xone \gets \tilde x_P\cdot(\Ztwo\cdot v_1(P) - \Zone\cdot v_0(P)) \)
    \tcp*{3\textbf{M} + 1\textbf{s}}
    \( \Tseven \gets \Zone\cdot \tilde y_P + \Ztwo\cdot \tilde x_P \)
    \tcp*{2\textbf{M} + 1\textbf{a}}
    % ...
    \( \Xtwo \gets \Tone\cdot v_1(P) + \Tseven\cdot v_0(P) \)
    \tcp*{2\textbf{M} + 1\textbf{a}}
    \( \Cfive \gets \Zone^2 - \Tseven\cdot \tilde x_Q \)
    \tcp*{1\textbf{M} + 1\textbf{S} + 1\textbf{s}}
    \( \Csix \gets \Tone \cdot \tilde x_Q + \Ttwo \)
    \tcp*{1\textbf{M} + 1\textbf{a}}
    \(
        \Tthirteen
        \gets 
        \tilde z_\oplus\cdot \tilde x_\ominus 
        - 
        \tilde x_\oplus\cdot \tilde z_\ominus
    \)
    \tcp*{2\textbf{M} + 1\textbf{s}}
    % ...
    \(
        \Xthree
        \gets 
        \Xone\cdot\Tthirteen
        -
        \Xtwo\cdot (
            \tilde x_\oplus\cdot \tilde y_\ominus 
            - 
            \tilde y_\oplus \cdot \tilde x_\ominus
        )
    \)
    \tcp*{4\textbf{M} + 2\textbf{s}}
    \( (\Xfive,\Xsix) \gets (\Xthree\cdot\Cfive, \Xthree\cdot\Csix) \)
    \tcp*{2\textbf{M}}
    \(
        \Xfour
        \gets 
        \Tthree
        \cdot(
            \Xone\cdot(
                \tilde z_\oplus\cdot \tilde y_\ominus 
                - 
                \tilde y_\oplus\cdot \tilde z_\ominus
            )
            + 
            \Tthirteen\cdot\Xtwo
        )
    \)
    \tcp*{5\textbf{M} + 1\textbf{a} + 1\textbf{s}}
    \( 
        (\Xseven,\Xeight)
        \gets 
        (\Xfive + \Zone\cdot\Xfour,
        \Xsix + \Ztwo\cdot\Zfour) 
    \)
    \tcp*{2\textbf{M} + 2\textbf{a}}
    \( 
        \Tsix 
        \gets 
        \tilde x_\oplus \cdot \tilde x_\ominus
    \)
    \tcp*{1\textbf{M}}
    % ...
    \(
        \E 
        \gets 
        - \Tsix
        \cdot 
        \Tthree 
        \cdot (
            \E\cdot\tilde x_P^2
            + 
            (\Xone\cdot \Tthree)^2
        )
    \)
    \tcp*{5\textbf{M} + 2\textbf{S} + 1\textbf{a} + 1\textbf{neg}}
    \( (\Xnine,\Xten) \gets (\E\cdot\Xseven,\E\cdot\Xeight) \)
    \tcp*{2\textbf{M}}
    \(
        \F 
        \gets
        \Xtwo\cdot(
            \tilde x_\oplus\cdot \tilde y_\ominus 
            + 
            \tilde y_\oplus\cdot \tilde x_\ominus
        ) 
        + 
        \Xone\cdot(
            \tilde z_\oplus\cdot \tilde x_\ominus 
            + 
            \tilde x_\oplus\cdot \tilde z_\ominus
        )
    \)
    \tcp*{6\textbf{M} + 3\textbf{a}}
    % ...
    \(
        \F 
        \gets 
        \Xone\cdot\F
        +
        2(\Xtwo^2\cdot \Tsix)
    \)
    \tcp*{2\textbf{M} + 1\textbf{S} + 2\textbf{a}}
    % ...
    \(
        \F 
        \gets
        -2(
            \F      
            \cdot \D
            \cdot \Tsix
            \cdot \Tthree
            \cdot \Tthree^2
            \cdot \tilde x_P
        )
    \)
    \tcp*{5\textbf{M} + 1\textbf{S} + 1\textbf{a} + 1\textbf{neg}}
    \(
        (\Uone,\Uzero)
        \gets 
        (- \F\cdot \tilde y_Q, \F\cdot \tilde z_Q)
    \)
    \tcp*{2\textbf{M} + 1\textbf{neg}}
    % ...
    \( \Fi \gets 1/(\F\cdot\tilde x_Q) \)
    \tcp*{1\textbf{M} + 1\textbf{I}}
    % ...
    \(
        (u_1',u_0',v_1',v_0')
        \gets 
        (\Fi\cdot\Uone,
         \Fi\cdot\Uzero,
         \Fi\cdot\Xnine,
         \Fi\cdot\Xten)
    \)
    \tcp*{4\textbf{M}}
    \Return{ \(\Mumford{X^2 + u_1'X + u_0'}{v_1'X + v_0'}\) }
\end{algorithm}

\subsection{Full scalar multiplication on \(\Jac{\C}\)}

We now combine our pseudo-scalar multiplication 
function \texttt{crypto\_scalarmult}
with the point-recovery function \texttt{Recover}
to define 
a full scalar multiplication function
\texttt{jacobian\_scalarmult} (Algorithm~\ref{alg:jacobian_scalarmult})
on \(\Jac{\C}\).

\begin{algorithm}
    \caption{\texttt{jacobian\_scalarmult}:
        Scalar multiplication on \(\Jac{\C}\),
        using the Montgomery ladder on \(\Kum{\C}\) 
        and recovery to \(\Jac{\C}\).
        Assumes wrapped projected point as auxiliary input.
        Symbolic cost:
        \(
            (7\beta+143)\textbf{M} 
            +
            (12\beta+12)\textbf{S}
            +
            (12\beta+4)\mathbf{m_c}
            + 
            (70+16\beta)\textbf{a}
            +
            (22+16\beta)\textbf{s}
            +
            3\textbf{neg} 
            +
            \textbf{I} 
        \).
        ATmega:
        {\avrjacscal} cycles.
        Cortex:
        {\armjacscal} cycles.
    }
    \label{alg:jacobian_scalarmult}
    \SetKwFunction{Project}{Project}
    \SetKwFunction{xWRAP}{xWRAP}
    \SetKwFunction{xUNWRAP}{xUNWRAP}
    \SetKwFunction{Recover}{Recover}
    \SetKwData{xP}{xP}
    \SetKwData{w}{w}
    \SetKwData{X}{X}
    \SetKwData{R}{R}
    \KwIn{\( (m,P,(x_P/y_P,x_P/z_P,x_P/t_P)) \in [0,2^\beta)\times\Jac{\C} \) }
    \KwOut{\( [m]P \in \Jac{\C} \)}
    \BlankLine
    \(
        (\X_0,\X_1) 
        \gets
        \texttt{crypto\_scalarmult}(m,(x_P/y_P,x_P/z_P,x_P/t_P))
    \)
    \;
    \tcp*{\((7\beta+4)\)M+12\(\beta\)S+12\(\beta m_c\)+16\(\beta\)a+16\(\beta\)s}
    \( \xP \gets \xUNWRAP((x_P/y_P,x_P/z_P,x_P/t_P)) \)
    \tcp*{4M}
    \Return{\(\Recover(P,\xP,\X_0,\X_1) \)}
    \tcp*{139M+12S+4\(m_c\)+70a+22s+3neg+1I}
\end{algorithm}

\begin{remark}
We have designed \texttt{jacobian\_scalarmult}
to take not only a scalar~\(m\) and a Jacobian point \(P\)
in its Mumford representation,
but also the wrapped form of \(\pm P\) as an auxiliary argument:
that is, we assume that \(\texttt{xP} \gets \texttt{Project}(P)\)
and \(\texttt{xWRAP}(\texttt{xP})\) have already been carried out as a
precomputation.
This saves redundant \texttt{Project}ing and \texttt{xWRAP}ping 
when we are operating on fixed base points, 
as is often the case in our protocols.
Nevertheless,
\texttt{jacobian\_scalarmult} 
could easily be converted to a ``pure'' Jacobian scalar multiplication function
(with no auxiliary input)
by inserting appropriate \texttt{Project} and \texttt{xWRAP}
calls at the start, and removing the \texttt{xUNWRAP} call at Line~2.
These modifications would increase the cost of \texttt{jacobian\_scalarmult}
by 11\textbf{M} + 1\textbf{S} + 4\(\mathbf{m_c}\) + 7\textbf{a} +
8\textbf{s} + 1\textbf{I}.
\end{remark}

%
%
%

\def\armcycles{$3\,589\,850$\xspace}
\def\armbytes{$7\,900$\xspace}
\def\armstack{$548$\xspace}

\def\avrfastcycles{$13\,900\,397$\xspace}
\def\avrfastbytes{$17\,710$\xspace}
\def\avrfaststack{$494$\xspace}

\def\avrsmallcycles{$14\,146\,844$\xspace}
\def\avrsmallbytes{$9\,912$\xspace}
\def\avrsmallstack{$510$\xspace}

\section{Results and comparison}
\label{sec:results}

The high-level cryptographic functions for our signature scheme
are named {\tt keygen}, {\tt sign} and {\tt verify}. 
Their implementations contain no surprises: they do exactly what
what was specified in~\S\ref{sec:highsig}, calling the lower-level
functions described in~\S\ref{sec:building-blocks}
and~\S\ref{sec:scalar-mult} as required. 
Our key exchange uses only the function {\tt dh\_exchange}, 
for both Diffie-Hellman key generation and key exchange. 
It implements exactly what we specified in~\S\ref{sec:highdh}: it is a call to {\tt crypto\_scalarmult} plus a call to {\tt xWRAP} to convert to the correct 384-bit representation. 
Table~\ref{tab:signaturesdh} (in the introduction)
presents the cycle counts and stack usage for all of our high-level functions.

\subsection{Code and compilation}

In our experiments,
the code for the AVR ATmega was compiled with {\tt avr-gcc} 
at optimization level {\tt -O2}. 
The ARM Cortex M0 code uses the {\tt clang} compiler, 
also with optimization level {\tt -O2}. 
We experimented with different optimization levels ({\tt -O3}, {\tt -O1}, and {\tt -Os}),
but the results were fairly similar. 
The total size of the program is \avrrom bytes for the AVR ATmega,
and \armrom bytes for the ARM Cortex M0. 
This consists of the full signature and key-exchange code, including 
the hash function {\tt SHAKE128}  
implemented with fixed 512-bit
output, with the code taken from 
the reference implementation.\footnote{We used 
    the reference C implementation for the Cortex M0,
    and the assembly implementation for AVR;
    both are available
    on \href{http://keccak.noekeon.org/}{\texttt{http://keccak.noekeon.org/}}.
    The only change required is to the padding,
    which must take domain separation into account
    according to~\cite[p.28]{FIPS202}.
}

\subsection{Comparison}

As we believe ours to be the first genus-2 hyperelliptic curve
implementation on both the AVR ATmega and the ARM Cortex M0
architectures, it is difficult to make a comparison.
On the other hand, we may compare with elliptic curve-based
alternatives at the same 128-bit security level:
notably~\cite{LWG14}, \cite{HS13}, \cite{WUW13}, and~\cite{DHH+15}.

If one is only interested in a key exchange scheme, 
it is enough to have an object which merely has a differential addition. 
A well-known example is the Montgomery model for elliptic curves,
with very efficient $x$-coordinate-only arithmetic,
which was used in eg.~\cite{LWG14}, \cite{HS13}, and~\cite{DHH+15} 
to obtain efficient Diffie--Hellman key exchange. 
It is also possible to use $x$-only arithmetic on Weierstrass curves,
as we see in~\cite{WUW13} 
(which is based on the pseudo-addition presented in~\cite{Brier--Joye}).
In genus 2, Kummer surfaces have similar properties.
Although they only have a pseudo-addition, it is very efficient and
therefore highly suitable for key exchange.

If one also wants to implement signatures, differential addition 
is no longer sufficient. In this case, one has to work with points
in an elliptic curve group (or the Jacobian of a genus 2 curve). 
To still make use of the efficient $x$-only arithmetic, 
one must project a point $(x,y)$ to the $x$-only representation, 
do a pseudo-scalar multiplication, and then recover the correct
$y$-coordinate of the result. 
(This is done in ~\cite{WUW13} using a Weierstrass curve.)
As recovery is generally quite slow, this imposes non-negligible overhead. 

In genus 2 we can use a completely analogous technique: 
we project points from the Jacobian to the Kummer surface,
use its efficient arithmetic for the scalar multiplication,
and recover the correct element of the Jacobian.
As in the elliptic case, this does impose some overhead. 
We therefore only do this when really necessary: that is, for signatures.
When computing shared secrets in key exchange, we remain on the Kummer surface.

\begin{table}[h!]
\centering
\renewcommand{\tabcolsep}{0.1cm}
\renewcommand{\arraystretch}{1.1}
	\begin{tabular}{|r|l|c|r|r|r|}
\hline
& {\bf Implementation} & {\bf Object} & {\bf Clock cycles} & {\bf Code size} & {\bf Stack} \\
\hline
{\bf S},{\bf DH} & Wenger et al.~\cite{WUW13} & NIST P-256 & $\approx 10\,730\,000 $& $7\,168$ bytes & $540$ bytes \\
\hline
{\bf DH} & D{\"u}ll et al.~\cite{DHH+15} & Curve25519 & \armcycles & \armbytes bytes & \armstack bytes \\
\hline
{\bf DH} & {\bf This work} & \(\Kum{\C}\) & \armscal & $\approx$~\armscalrom bytes & \armscalram bytes \\
\hline
{\bf S} &{\bf This work} & \(\Jac{\C}\) & \armjacscal & $\approx$~\armjacscalrom bytes & \armjacscalram bytes \\
\hline
\end{tabular}
\vspace{0.2cm}
	\caption{Comparison of scalar multiplication routines on the ARM
        Cortex M0 architecture at the 128-bit security level.
        {\bf S} denotes signature-compatible full scalar multiplication;
        {\bf DH} denotes Diffie--Hellman pseudo-scalar multiplication.}
	\label{tab:m0res}
\end{table}

As we see in Table~\ref{tab:m0res}, 
genus-2 techniques give great results 
for Diffie--Hellman key exchange on the ARM Cortex M0 architecture.
Comparing with the current fastest implementation~\cite{DHH+15}, we reduce
the number of clock cycles by about $27\%$, while about halving code
size and stack usage. For signatures, the state-of-the-art
is~\cite{WUW13}: here we reduce the cycle count for the 
underlying scalar multiplications by a very impressive $75\%$,
at the cost of a moderate increase in code size and stack usage.

\begin{table}[h!]
\centering
\renewcommand{\tabcolsep}{0.1cm}
\renewcommand{\arraystretch}{1.1}
	\begin{tabular}{|r|l|c|r|r|r|}
\hline
& {\bf Implementation} & {\bf Object} & {\bf Cycles} & {\bf Code size} & {\bf Stack} \\
\hline
    {\bf DH}    & Liu et al.~\cite{LWG14}                    & $256$-bit
curve& $\approx 21\,078\,200$ & $14\,700$ bytes$^{*}$         &  $556$ bytes         \\
    \hline
    {\bf S,DH}  & Wenger et al.~\cite{WUW13}              & NIST P-256           & $\approx 34\,930\,000$ & $16\,112$ bytes                  &  $590$ bytes         \\
    \hline
    {\bf DH}    & Hutter, Schwabe~\cite{HS13}                             & Curve25519           & $22\,791\,579$         & n/a$^{\dagger}$                     &  $677$ bytes         \\
    \hline
    {\bf DH}    & D{\"u}ll et al.~\cite{DHH+15} & Curve25519 & \avrfastcycles & \avrfastbytes bytes & \avrfaststack bytes \\
    \hline
    {\bf DH}    & {\bf This work} & \(\Kum{\C}\) & \avrscal & $\approx$~\avrscalrom bytes & \avrscalram bytes \\
\hline
    {\bf S}     & {\bf This work} & \(\Jac{\C}\) & \avrjacscal & $\approx$~\avrjacscalrom bytes & \avrjacscalram bytes \\
\hline
\end{tabular}
\vspace{0.2cm}
	\caption{Comparison of scalar multiplication routines 
        on the AVR ATmega architecture
        at the 128-bit security level.
        {\bf S} denotes signature-compatible full scalar multiplication;
        {\bf DH} denotes Diffie--Hellman pseudo-scalar multiplication.
        The implementation marked $^{*}$ also contains a fixed-basepoint
        scalar multiplication routine, whereas the implementation marked
        $^{\dagger}$ does not report code size for the separated scalar
        multiplication.}
	\label{tab:avrres}
\end{table}

Looking at Table~\ref{tab:avrres},
on the AVR ATmega architecture we reduce the cycle count for
Diffie--Hellman by about $32\%$ compared with the current
record~\cite{DHH+15}, again roughly halving the code size, 
and reducing stack usage by about $80\%$. 
The Jacobian scalar multiplication needed for signatures,
reduces the cycle count by $71\%$ compared to~\cite{WUW13}, 
while increasing the stack usage by $25\%$.


Finally we can compare to the currently fastest full signature implementation~\cite{NLD15},
shown in Table~\ref{tab:sigres}.

\begin{table}[h!]
\centering
\renewcommand{\tabcolsep}{0.1cm}
\renewcommand{\arraystretch}{1.1}
	\begin{tabular}{|l|c|r|r|r|}
\hline
{\bf Implementation} & {\bf Object} & {\bf Function} & {\bf Cycles} & {\bf Stack} \\
\hline
    \hline
    Nascimento et al.~\cite{NLD15} & Ed25519 & sig. gen. & $19\,047\,706$ & $1\,473$ bytes \\
    \hline
    Nascimento et al.~\cite{NLD15} & Ed25519 & sig. ver. & $30\,776\,942$ & $1\,226$ bytes \\
    \hline
    {\bf This work} & \(\Jac{\C}\) & {\tt sign} & \avrsign & \avrsignram bytes \\
\hline
    {\bf This work} & \(\Jac{\C}\) & {\tt verify} & \avrverify & \avrverifyram bytes \\
\hline
\end{tabular}
\vspace{0.2cm}
	\caption{Comparison of a full signature scheme 
        on the AVR ATmega architecture
        at the 128-bit security level.
        }
	\label{tab:sigres}
\end{table}
We see that we almost half the number of cycles, while also reducing the stack usage by a 
decent margin.
We do not compare code size, as this is not reported in \cite{NLD15}.

\bibliographystyle{plain}
\bibliography{collection}


\end{document}